\documentclass[aip,jap,reprint]{revtex4-1}
\usepackage{graphicx}                 
\usepackage{dcolumn}                
\usepackage{bm}                        
\usepackage[mathlines]{lineno}     
\usepackage[colorlinks]{hyperref}
\newcommand{\Heb}{H_{eb}}
\newcommand{\Hc}{H_c}
\newcommand{\Ms}{M_s}
\newcommand{\cites}[1]{\cite{#1}}  
\newcommand{\refs}[1]{\ref{#1}}    

\begin{document}
\title{Coercivity and random interfacial exchange coupling in CoPt/Co films} 
\author{V.~Alexandrakis}
\affiliation{Institute of Nanoscience and Nanotechnology, NCSR Demokritos, 15310 Athens, Greece}
\author{D.~Kechrakos}
\email[Corresponding author:]{ dkehrakos@aspete.gr}
\affiliation{Department of Education, School of Pedagogical and Technological Education, 14121 Athens, Greece}
\author{N.~Moutis}
\affiliation{Department of Education, School of Pedagogical and Technological Education, 14121 Athens, Greece}
\author{D.~Niarchos}
\affiliation{Instiute of Nanoscience and Nanotechnology, NCSR Demokritos, 15310 Athens, Greece}
\author{G.~Hadjipanayis}
\affiliation{Department of Physics and Astronomy, University of Delaware, Newark, 19716 Delaware , USA}
\author{I.~Panagiotopoulos}
\affiliation{Department of Materials Science and Engineering, University of Ioannina, 45110 Ioannina, Greece}
\date{\today}

\begin{abstract}
Hard-soft bilayers are analogous to prototype exchange-biased ferromagnetic -antiferromagnetic systems as the minor loop of the soft layer is biased by the hard and furthermore they offer bias layer tunability. In sputtered CoPt/Co hard-soft bilayers we demonstrate that the exchange bias field shows a linear dependence on the hard layer magnetization, while the coercivity shows a quadratic dependence. Analysis of the minor hysteresis loop features supported by Monte-Carlo simulations provide clear evidence that the coercivity of the soft layer is mainly determined by the tunable randomness of the domain state of the hard layer. 
\end{abstract}
\maketitle    

\section{Introduction}
Exchange bias phenomena in layered structures\cites{nog99,hel07} and nanostructures\cites{nog05,ort15} attract a great deal of research effort related to fundamental issues regarding the underlying magnetization reversal mechanism and also their numerous technological applications ranging from biomedicine to magnetic recording industry. 
Among the fundamental issues we mention the increase in coercivity that most commonly accompanies exchange biasing (EB) phenomena in antiferromagnetic (AF) - Ferromagnetic (FM) systems\cites{nog99, nog05} .
For instance, chemically synthesized Ni/NiO nanocomposites show both large exchange bias and enhanced coercivity.\cites{yao14} 
In nanopatterned Co/CoO island structures a strong increase in bias and coercive field were found as the nanostructure size was reduced only for thickness below 12nm.\cites{lau12} 
In (Pt/Co)$_3$ multilayer stacks it was found that the coercivity depends sensitively on structural changes as subtle as a sub-monolayer interface oxidation of a ferromagnetic layer.\cites{kos12}
Recently it was reported that in the vicinity of the compensation composition in Mn–Pt–Ga ferrimagnetic Heusler alloy, a giant exchange bias field of more than 3~T and a large coercivity are established.\cites{nay15}
Monte Carlo studies of exchange bias in FM/AF bilayers\cites{jia15} 
show a weak increase of coercivity when the interface exchange coupling is stronger than the anisotropy of the hard AF layer. 
Interestingly, in many layered systems the coercivity of the FM layer as a function of temperature shows a peak at the blocking temperature, while as a function of the AF layer thickness shows a maximum at the critical thickness below which EB disappears. 
Within the large diversity of systems that exhibit EB several aspects of these phenomena have been explained using the random interfacial field model.\cites{mal87}
The random interfacial exchange coupling can lead to the formation of AF domains\cites{xi03} and furthermore, during magnetization reversal, can break the ferromagnetic layer into domains smaller than the ones occurring without the coupling to the antiferromagnetic layer.\cites{lei00}
On the other hand, formation of volume AF domains\cites{now02,lee06} that can be controlled by diluting the antiferromagnetic layer with nonmagnetic substitutions\cites{kel02} has been used to explain a variety of typical effects associated with exchange bias. 
It is reasonable to expect that in the presence of random fields at an interface between a FM and an AF layer, the domain walls in the FM layer are pinned at local energy minima\cites{zha99} and that for the domain walls to move, the applied magnetic field must be large enough to overcome the statistical fluctuations of energy. 
Random pinning effects have been studied in Cu/Co spin valves biased by thermally oxidized NiO layers with increased roughness\cites{chi07}.
Studies of the coupling between hard FePt/CoFe  and soft CoFe/NiFe bilayers through a Cu interlayer show  that the bias does not depend linearly on the percentage of hard magnetic layer switching but still the coercivity exhibits a maximum at the point of zero bias, which takes place at approximately 70\% switching of the hard layer for the 3.5nm Cu buffer.\cites{zha09} 
In  MnF$_2$/Fe bilayers it is found\cites{lei00} that when the antiferromagnetic surface is in a state of maximum magnetic frustration and the net exchange bias is zero, there is  strong enhancement of the coercivity, which is proportional to the exchange coupling between the layers.
The large number of different systems that exhibit EB, in combination with the lack of exact information on the interfacial spin structure in each case, are two main limiting factors in establishing a global theory of the exchange bias effect. 
Hard FM-Soft FM bilayer systems on the other hand, mostly in the weak coupling limit, have some resemblance with the archetype AF-FM exchange bias systems\cites{dum00,ful99,bin06,man98} in the sense that the minor hysteresis loop of the soft layer is biased by the hard. 
Thus, much of the physics and reversal mechanisms are expected to be similar if the AF layer is replaced by a hard FM as the biasing layer.  
At the same time hard-soft (HS) systems are simpler to understand since the magnetic state of the biasing layer is more straightforward to probe and adjust.\cites{ber04,ber05} 
In addition the HS FM bilayers offer bias tunability \cites{ber04,ber05}, that allows control over the loop shift and width through the field-dependence of the magnetic state of the hard layer.

Despite the anticipated less complex nature of the FM HS bilayers, conflicting trends on the correlation between the exchange bias field and the coercivity of Co-based HS system have been reported. 
Namely, although in both CoPtCrB/Co and CoPt/Co HS bilayers the exchange bias field shows a linear dependence on the hard layer magnetization, 
increasing coercivity with bias (hard) layer  magnetization is reported for the CoPtB/Co system\cites{ber04,ber05} and the opposite behavior is reported for the CoPt/Co bilayer\cites{kte08}. 
This point motivated further investigation of the EB behavior of  CoPt/Co bilayers and the underlying magnetization reversal mechanism that is reported here. 
In the present work, we study the CoPt/Co sputtered bilayers by standard magnetometry combined with Monte Carlo simulations that support the experimental findings and shed light on the underlying magnetization reversal mechanism. It is shown that in CoPt/Co sputtered bilayers the EB field of the soft layer varies linearly with the hard layer magnetization, while the sample coercivity depends quadratically on the exchange bias field. 
Our numerical simulations support the domain wall displacement mechanism as the dominant mechanism of magnetization reversal, in contrast to the case of CoPtCrB/Co bilayers where an exchange-correlated coherent reversal mechanism was put forward.\cites{ber04}
\section{Experimental Details }
The Co$_{45}$Pt$_{55}$ films were deposited by magnetron sputtering at ambient temperature on oxidized Si (001) substrates at a rate of 1.43~\AA/s, in an Ar gas pressure of 3~mTorr. 
The as-deposited films are magnetically soft and their XRD patterns indicate a fcc structure with (111) texture.
In order to produce the bilayered hard-soft structure, after deposition of the bottom (hard) CoPt layer of thickness $t_{HL}$=20 nm the films were annealed in high vacuum ( $<2 \times 10^{-6}$ Torr ) in order to crystallize the high anisotropy L1$_0$ phase. 
Then the sample was cooled to room temperature and the top (soft) Co layer of thickness $t_{SL}$=8 nm was deposited.  
The layer thicknesses were determined by X-ray reflectivity. 
The magnetic measurements were performed with a Lake Shore vibrating sample magnetometer (VSM) and in-plane magnetic field. 
The Magnetic Force Microscopy (MFM) images have been obtained with the use of a NT-MDT scanning probe microscope in semi-contact mode.
\begin{figure} [htb!]    
\includegraphics[scale=0.60]{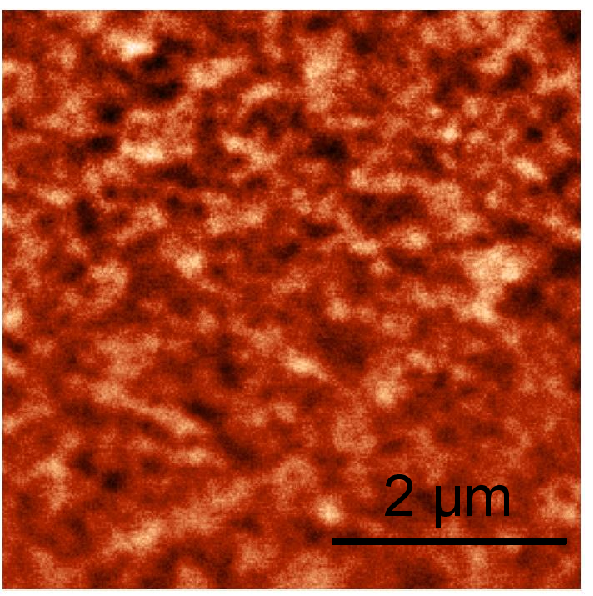}
\includegraphics[scale=0.60]{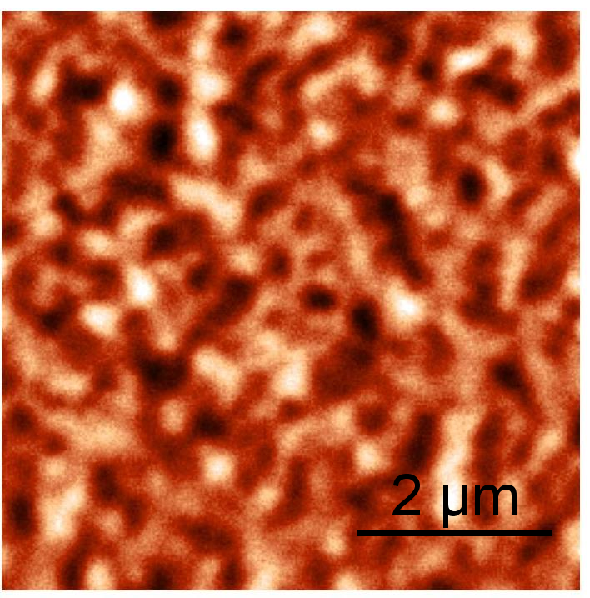}
\caption{Magnetic force microscopy images of CoPt/Co bilayer at different magnetic states of the hard layer.  Remanent state (left) of a sample area $5\times 5~\mu$m$^2$ and  demagnetized state (right) of a sample area $6\times 6~\mu$m$^2$ are shown.}
\label{exp_mfm}
\end{figure}
\begin{figure}  [htb!]    
\includegraphics[scale=1.00]{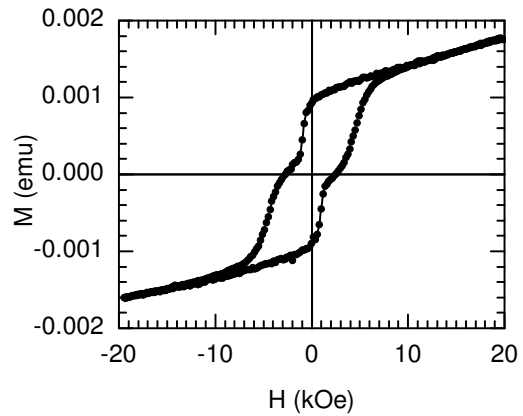}
\caption{Full hysteresis loop of CoPt/Co film.}
\label{exp_full}
\end{figure}
\begin{figure}  [htb!]    
\includegraphics[scale=1.00]{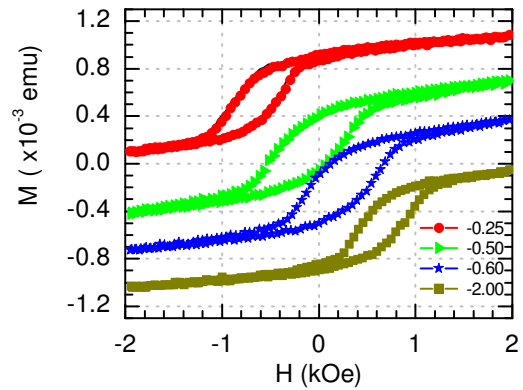}
\caption{Typical set of minor hysteresis loops measured at different magnetic states of the hard layer as indicated by the different values of the applied field -0.25, -0.50, -0.60 and -2.00~kOe}
\label{exp_loops}
\end{figure}
\begin{figure} [htb!]    
\includegraphics[scale=1.10]{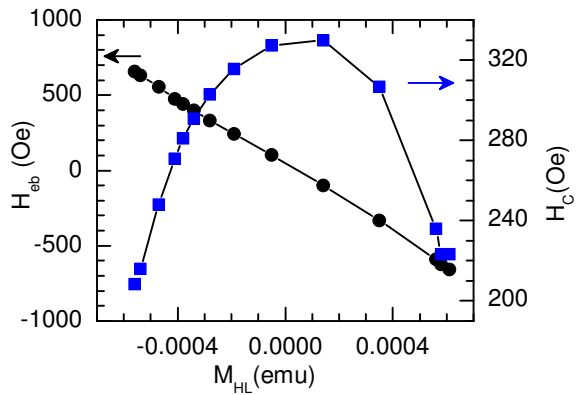}
\caption{Linear dependence of exchange-bias field (left axis) and quadratic dependence of coercivity (right axis) on the hard-layer magnetization.} 
\label{exp_hchbmr}
\end{figure}

\section{Experimental Results }

The MFM images of Fig.\refs{exp_mfm} depict the remanent and DC-demagnetized states of a single CoPt hard layer. 
They represent areas of $5\times 5~\mu$m$^2$ and $6\times 6~\mu$m$^2$, respectively. 
The demagnetized state is characterized by strong contrast maze-like domain patterns of average width $0.25~\mu$m and length $1~\mu$m. 
The remanent state has thinner reversed stripes (typically $0.1~\mu$m) with smeared out contrast. 
To asses the effect of the magnetic state on the reversal of the soft phase, minor loops that characterize the soft layer reversal have been monitored for different magnetic states of the hard layer.
This is feasible due to the weak coupling between the two phases evidenced by the fact that the full hysteresis loop shows a well defined shoulder between the switching of the two phases (Fig.\refs{exp_full}). 
Thus the switching field distributions of the hard and the soft phases are well separated and the minor loops are well defined: a field of  $\sim$2~kOe fully saturates the soft phase and yields a closed loop\cites{ges08}, but it does not affect significantly the hard one. 

Families of minor hysteresis loops have been monitored for different magnetization states of the hard layer.
In particular, the bilayer  sample was first brought to positive saturation by applying +20kOe. 
Then a reverse field $H_{demag}$ in the range -20kOe$\le H_{demag} \le$ -3kOe was applied,
which drives the hard layer to a partially demagnetized state.
A typical set of measurements is shown in Fig.\refs{exp_loops}. 
As expected the minor loops are not centered in the origin of the M-H plots: 
The vertical displacement $M_H$ (along the magnetization axis) represents the contribution of the hard layer, that is, the part of the magnetic moment that remains unswitched during the sweep of the mirror loop. 
The loop displacement along the field axis  represents the exchange bias field ($\Heb$)and the soft layer coercivity $H_c$ must be defined as the halfwidth of the loop.  
The squareness of the minor loop can be also defined analogously, taking of course on account the loop shift, as $S= (\Ms(\Heb)-M_H)/\Ms(2$~kOe).  
The observed $S$ values are consistent with the $2/\pi$ value expected for a random in-plane easy axes distribution.
An obvious observation is the linear relationship between $\Heb$ and $M_H$ (Fig.\refs{exp_hchbmr}), which shows that the interfacial coupling can be described by a simple linear term, $-JM_H \cdot \Ms$. 
That means that the interfacial ordering of the magnetic moments can be assumed to be represented by the magnetic state of the whole hard layer. 
This is  plausible given that the exchange field ($\approx$ 650~Oe) is much lower than the coercivity of the hard layer ($\approx$ 5400~Oe) (and its anisotropy field therefore ) and that the domains of a soft phase are expected to be much larger.
The data can be described by a linear fit 
$\Heb$(Oe) = $675(\pm 5)\cdot M_H +  32(\pm 3)$.  
The coercivity as a function of the vertical shift, and consequently $\Heb$, shows a very simple quadratic dependence, which is more symmetric if we choose to plot $\Hc$ versus $\Heb$. 
The data can be described as:
$\Hc$(Oe)= $335(\pm 2)\cdot (1-(\Heb/1100)^2)$
This simple quadratic dependence is indicative of a possible link of the coercivity to the randomness of the magnetic state through the random interfacial exchange coupling. 
Simply put, for an assembly of randomly oriented vectors of constant length say $M_S$ the standard deviation is expected to be $\sigma^2= <M^2>-<M>^2=Ms^2-M^2$.

\section{Numerical Modeling }
\begin{figure} [htb!]    
\includegraphics[scale=0.45]{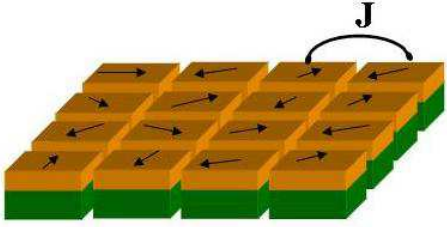}
\caption{Model of the CoPt/Co bilayer composed of identical grains. The lower part of each grain is the hard material and the upper part is the soft material.  Arrows indicate the direction of the random exchange bias field ($H_R$) acting on the soft part of each grain. Exchange coupling ($J$) is assumed between neighboring grains.}
 \label{fig_film}
 \end{figure}

To gain further insight into the magnetization reversal mechanism leading to the experimentally observed characteristics of the (minor) hysteresis loops we have conducted a series of Monte Carlo simulations of magnetic hysteresis of a simplified system that shares the same major features with the sputtered samples.
In particular, we assume that the morphology of the sputtered sample, due to its granular nature, can be adequately described by a two-dimensional array of identical bi-magnetic grains composed of a soft part exchanged coupled to a hard part (Fig.~\refs{fig_film}). 
The magnetization of the soft part is assumed to reverse coherently under application of an external field, while that of the hard part remains frozen. 
The effect of the hard-soft coupling across the interface of each grain can be  therefore approximated by a local bias field ($H_R$) acting on the soft part of the grain. 
The direction of the local bias field acting on grain-$i$ is taken in-plane, making a random angle $\phi_i$ with the external field ($H$). 
Neighboring grains of the SL are exchange coupled by ferromagnetic exchange interactions with strength $J$
The total energy of the system then reads 
\begin{eqnarray}
E = -K_1 \sum_{i}  ( \widehat{S}_{i}\cdot \widehat{e}_{i} )^2 
       -H    \sum_{i} S_{ix} 
			 -H_R \sum_{i} \cos{\phi_i}     \nonumber \\
       -J     \sum_{ij} (\widehat{S}_{i} \cdot \widehat{S}_{j} ) 
\label{eqn1}
\end{eqnarray}
where $\hat{S}_i$ denotes the magnetization direction (spin) of grain $i$, 
$K_1$ the anisotropy energy per grain and $\hat{e}_i$ the random easy axis direction
$H$ is the applied field along the $x$-axis and 
$H_R$ the random exchange field that makes an angle $\phi_i$ with the $x$-axis. 
The $\phi$-distribution depends on the magnetization state of the hard layer and to the simplest approximation is given as\cites{ber05},
$D(\phi)=(1+m_r)/2\pi$ for $-\pi/2 < \phi \le +\pi/2 $ and
$D(\phi)=(1-m_r)/2\pi$ for $+\pi/2 < \phi \le +3\pi/2 $, 
where $0 \le m_r \le 1$ is the normalized remanent magnetization of the hard layer.
Since energy units in Eq.\refs{eqn1} are arbitrary, we scale all energy parameters entering Eq.\refs{eqn1} by the exchange coupling constant ($J=1$).
Hysteresis loops are simulated using the standard Metropolis Monte Carlo algorithm with single spin updates and $10^4$ initial Monte Carlo steps per spin (MCSS) for thermalization followed by $10^4$ MCSS for thermal averaging. Sampling over thermal disorder is performed every $\tau=10$ MCSS to suppress correlations between sampling points. A field step of $\Delta H=0.001$ is used.
The percentage of accepted spin moves is kept close to $50\%$ by adjusting the width of the spin moves.
An array containing $33 \times 33$ grains is used for the simulations and configurational averages over the quenched randomness of the local easy axes and the local bias field directions is performed using an assembly of $N_a=10$ samples.

\begin{figure} 
\includegraphics[scale=0.6]{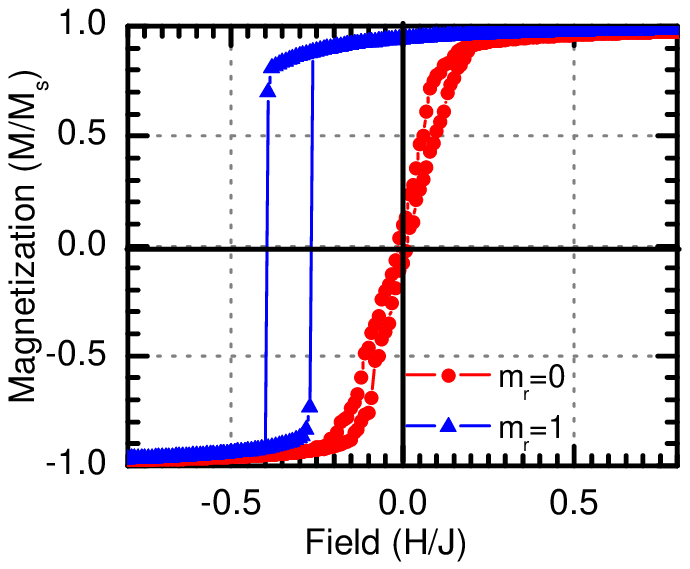}
\includegraphics[scale=0.6]{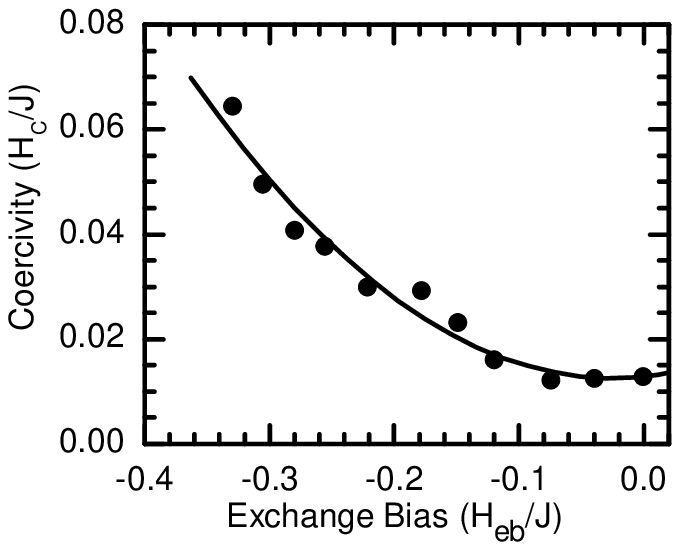}
\caption{Magnetic hysteresis of ferromagnetically coupled grains (left). Loops for an unbiased hard layer ($m_r=0$) and a hard layer at the remanent state ($m_r=1$) are shown and the resultant increase of coercivity with exchange bias field for different HL magnetization states (right). The solid line is a quadratic fit to the data points. Model parameters used are $K_1/J=0.5, H_R/J=0.5$ and temperature $T/J=0.001$}
 \label{hyst_mb}
 \end{figure}

In Fig.\refs{hyst_mb} we show results for the hysteresis behavior of a sample with exchange coupled bilayer grains. 
As previously measured  CoPtCrB-Co hard-soft bilayers\cites{ber04} and interpreted by statistical arguments\cites{ber05} the intergranular exchange introduces lateral correlations to the moments of the grains and leads eventually to increasing coercivity as the exchange bias increases (in absolute value). 
Furthermore, the observed increase of loop squareness\cites{ber04} as the HL magnetization approaches the saturation value $(m_r=1)$ is also reproduced in Fig.\refs{hyst_mb}. 
These findings support the assumption that the underlying magnetization reversal mechanism is the CoPtCrB-Co samples is weakly-correlated coherent rotation of the grain magnetizations.\cite{ber05} 

However, these findings contradict the experimental observations in the CoPt/Co bilayer studied in the present work. 
Thus we sought a different mechanism than that of the coherent rotation of the grain magnetization assumed in the previous analysis. 
The increase of the coercive field in the demagnetized state implies that a model of enhancement of $H_c$ due to random magnetic state inhomogeneities at the interface of the bilayer should be more suitable. 
This mechanism is justified by the fact that soft layer domains are much larger than those of the hard. 
Thus the soft layer domain walls propagate on a random pinning substrate defined by the partially (or fully) demagnetized hard layer. 
On the contrary, the mechanism of  Ref.~\cites{ber04,ber05} is expected to dominate only in the case that the soft layer grain of smaller size compared to the domain features of the hard layer. 
A full-scale modeling of the magnetization reversal mechanism via domain wall propagation would require consideration of the magnetostatic interactions in our system, a fact that would render the numerics rather time-demanding\cites{san06}. 

\begin{figure} 
\includegraphics[scale=0.15]{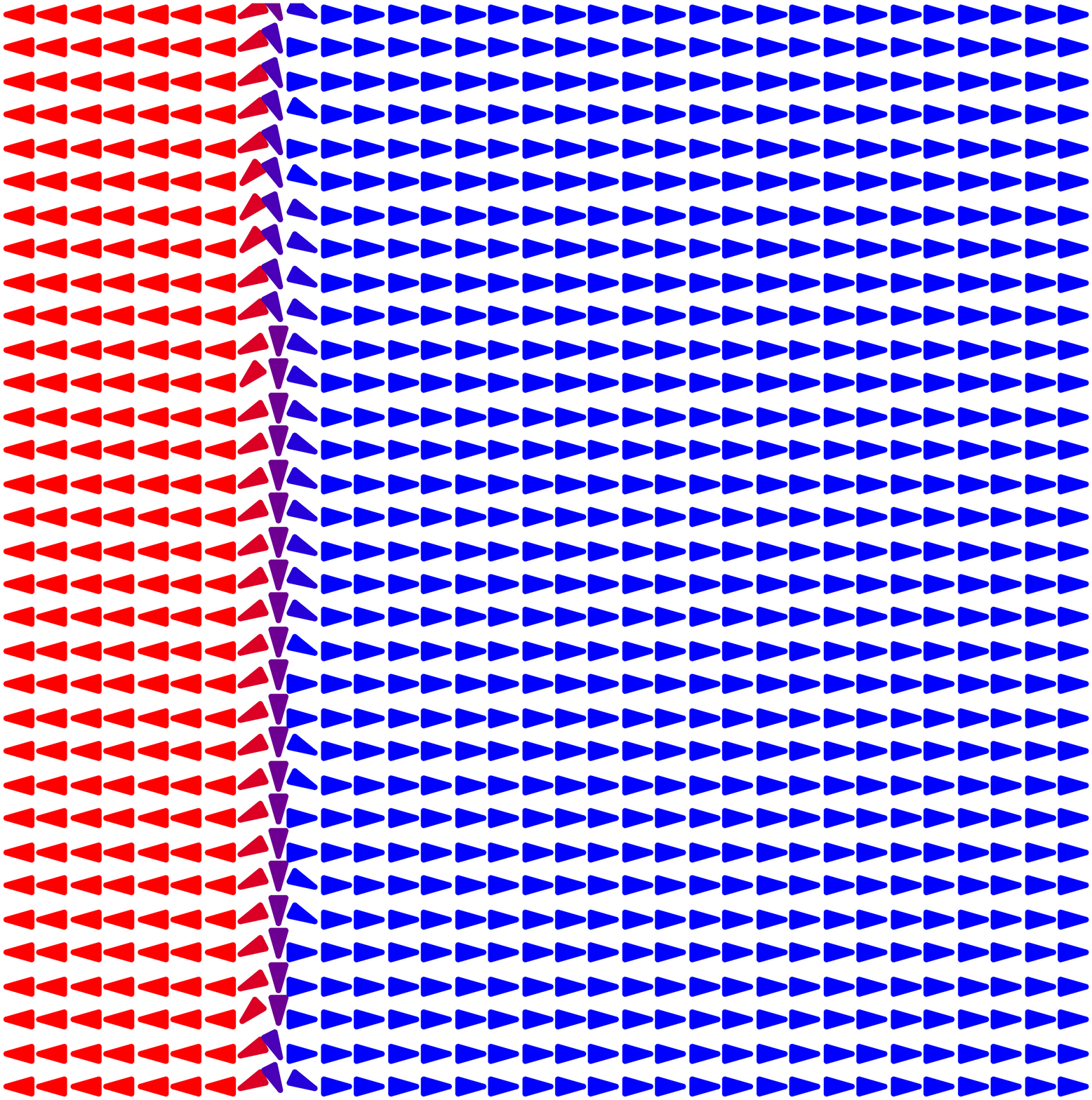}
\caption{Micromagnetic configuration of a sample at positive remanence ($H=0^+$) containing a thin domain wall ($\delta_w\approx 3$). Model parameters used are  $K_1/J=0.1, H_R/J=0.05$ and temperature $T/J=0.001$. The hard-layer is at the demagnetized state ($m_r=0$).}
 \label{dw_sketch}
 \end{figure}

\begin{figure} [htb!]    
\includegraphics[scale=0.6]{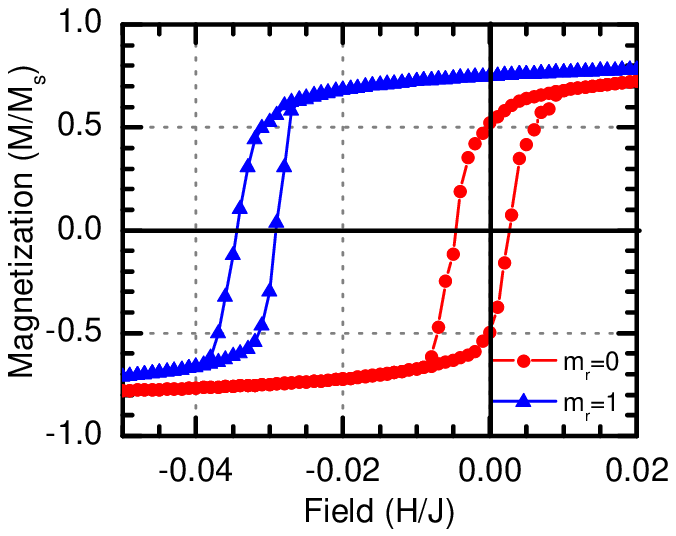}
\includegraphics[scale=0.6]{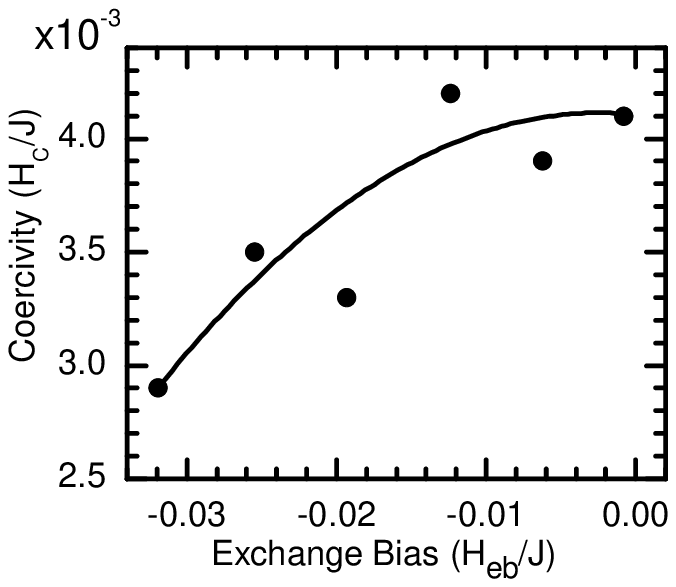}
\caption{Magnetic hysteresis due to domain wall motion (left) and the resultant quadratic dependence of coercivity on the exchange bias field (right). The motion of a Bloch wall with $\delta_w\approx 3$a along the x-axis is assumed (see Fig.\refs{dw_sketch}). The solid line is a quadratic fit to the data. Model parameters used are $K_1/J=0.5, H_R/J=0.05$ and temperature $T/J=0.001$}
 \label{hyst_dw}
 \end{figure}

We have thus adopted a simplified approach, that consists in generating a static domain wall in the sample, drive it via the external magnetic field and monitor the hysteresis loop changes for different magnetic states of the hard layer. 
To this end, we fix the magnetic moments on opposite sides of our sample ($x=0$ and $x=L$) to opposite directions along the $x$-axis thus generating a Bloch wall in the interior of our sample (Fig.\refs{dw_sketch}).
By cycling the applied field form positive to negative saturation we force the backward and forward  translation of the domain wall along the $x$-axis.
Sample parameters are chosen such that conditions of strong hindering of the domain wall by the magnetic roughness hold. 
This is achieved when the domain wall thickness $\delta_w \approx  \sqrt{J/K_1} $ is comparable to the length scale of roughness, which in our model is equal to the distance between grains ($\delta_r \approx 1$). 
As shown in Fig.~\refs{hyst_dw}, the loop width decreases as the remament state of the hard layer approaches the maximum value ($m_r=1$). 
The experimentally observed trend of the coercivity (Fig.\refs{exp_hchbmr}) is quite satisfactorily reproduced by the simulation data indicating that the degree of magnetic roughness of the hard layer determines the coercivity of the bilayer via hindering of the domain wall motion.

\section{Conclusions }
The easily tunable state of the magnetic state in weakly coupled hard-soft bilayers offers the possibility to understand by analogy the physics of $H_c$ in AF-FM exchange bias systems. 
In sputtered CoPt/Co hard-soft bilayers a simple linear dependence of the exchange bias field on the hard layer magnetization is found, while the cobalt layer coercivity shows a quadratic dependence on hard layer magnetization with the maximum obtained at the demagnetized state of the hard layer. 
By means of Monte Carlo simulations, we have examined two possible mechanisms which lead to opposite dependence of the coercivity on the exchange bias field. 
The experimental data are described by a model assuming domain wall pinning due to random magnetic state inhomogeneities at the interface of the hard-soft bilayer. 
The proposed mechanism is justified  by the fact that soft layer domains are much larger than those of the hard and have to propagate against the random pinning forces caused by the inhomogeneous magnetic state of the hard layer. 
This implies that the grains are much larger than the domain features of the hard layer and consequently sets the microstructural limits of validity of the model.

\begin{acknowledgments}
IP acknowledges the use of the VSM unit of the University of Ioannina Laboratory Network. This research was co-financed by the European Social Fund and Greek national funds through the Research Funding Program \small ARCHIMEDES-III (MIS 383576).
\end{acknowledgments}

\end{document}